# Polarized Negative Ion Source with Multiply Sphericaly Focusing Surface Plasma Ionizer

V. Dudnikov[1a], A. Dudnikov[2]

[1]Muons, Inc., Batavia, IL, USA, [2] Novosibirsk, Russia,

[a)]Corresponding Author:dvg43@yahoo.com

It is proposed one universal H-/D- ion source design combining the most advanced developments in the field of polarized ion sources to provide high-current high-brightness ion beams with >90% polarization and improved lifetime, reliability, and power efficiency. The new source utilizes high-efficiency resonant charge-exchange ionization of polarized neutral atoms by negative ions generated by cesiated surface-plasma interactions via a multi-spherical negative ion focusing element. Multi-spherical focusing of the negative ions strongly suppresses the parasitic generation of unpolarized H-/D- ions. By incorporating new and novel designs for the dissociator and plasma generator in parallel with the multi-spherical focusing the design can suppress adsorption and depolarization of particles from the polarized beam greatly improving performance over current concepts.

## INTRODUCTION

The beam parameters necessary to achieve the design luminosity of the polarized Medium-energy Electron-Ion Collider (MEIC) at JLab are shown in Ref. [1]. At MEIC, in addition to polarized protons, there is a particular nuclear physics interest to polarized d/He/Li and even heavier ion beams. The special design of the accelerators and storage rings (figure-8 shape) in this project should provide very good polarization preservation for all ion species including deuterons. It becomes possible to efficiently control the polarization of a beam of particles with any anomalous magnetic moment including particles with small anomalous moments, such as deuterons. This means that the final beam polarization after acceleration will be determined by the beam polarization extracted from the ion source, which must be made as high as possible. Availability of polarized deuterons and of other polarized light and even heavy ions opens new physics opportunities and is particularly important for maximizing the discovery potential of the electron-ion collider.

The evolution of polarized ion sources has been presented recently by W. Haeberli [2]. The most advanced versions of polarized H- sources are:

1. Atomic Beam Polarized Ion Source (ABPIS) with resonant charge-exchange ionization of polarized atoms by negative ions, developed and presented by A. Belov [3];
2. Optically-Pumped Polarized H- Sources (OPPIS) with Rubidium and Sodium targets, developed and presented by A. Zelenski [4].

The BNL OPPIS used in RHIC operation was upgraded recently [4] and can provide at injection into the RFQ up to 0.5 mA of polarized H- in 0.5 ms pulses with polarization of up to P=0.83. In this ion source, fast protons (2.5-3.5 keV) produced by high brightness arc discharge source are converted to fast atoms, injected into a strong magnetic field, converted to proton in He target, to pick up polarized electrons from optically pumped Rb vapor, pass through a Sona-Transition region, and pick up a second (unpolarized) electron from Na vapor forming a beam of nuclear-polarized H- ions. To prevent depolarization in the charge-exchange collisions, the optically pumped cell is located inside a strong (25-30 kG) superconducting solenoid.

Since the experimental efficiency is proportional to $P^2$, having the highest possible polarization is very important Therefore, it is difficult to use OPPIS for production of highly polarized D- ions.

In our opinion, the ABPIS [3, 5] has a higher potential for efficient production of different negative H-/D-/T- ions with polarization greater than 95%. It can be less expensive to manufacture and operate and is therefore more attractive as a commercial product.

# STATUS OF ABPIS

The most advanced version of the Atomic Beam Polarized Ion Source (ABPIS) with ionization of polarized atoms by resonant charge exchange with negative ions was developed at the Institute of Nuclear Research (INR RAN) in Troitsk, Russia [3,5]. A schematic diagram of this ABPIS is shown in Fig. 1. Recently an ABPIS of polarized protons was reproduced for the Dubna accelerator system [6].

A polarized atomic H$^-$ beam source with selective resonant charge-exchange ionization has a good potential to produce H$^-$/D$^-$ ion beams with the highest polarization [3,5,6,7,8,9]. Low energy unpolarized H$^-$/D$^-$ ions can transfer electrons only to H or D atoms but not to molecules. This method of ionization allows one to produce H$^-$/D$^-$ ion beam polarization higher than that of the atomic beam by eliminating unpolarized molecules from the beam. A pulsed mode of operation is favorable for high-intensity and high-polarization production. An efficient pulsed operation can be attained using a fast gas valve and a small-volume RF discharge dissociator with a helicon antenna in a magnetic field and AlN chamber for efficient cryogenic cooling (a new design of the dissociator).

In the ABPIS, hydrogen or deuterium atoms are formed by the dissociation of molecular gas, typically in an RF discharge dissociator. The atomic flux is cooled to a temperature of 30K - 80K by passing through a cryogenically cooled nozzle. The atoms escape from the nozzle orifice into the vacuum and are collimated to form a beam. The beam passes through a region with inhomogeneous magnetic field created by sextupole magnets where atoms with one orientation of the electron spin relative to the magnetic field are focused while atoms with the opposite orientation of the electron spin are defocused. Nuclear polarization of the beam is increased by inducing transitions between the spin states of the atoms. The RF transition units are also used for fast reversal of the nuclear spin direction without changing the atomic beam intensity and divergence. The main components of this ABPIS shown in Fig. 1 are:

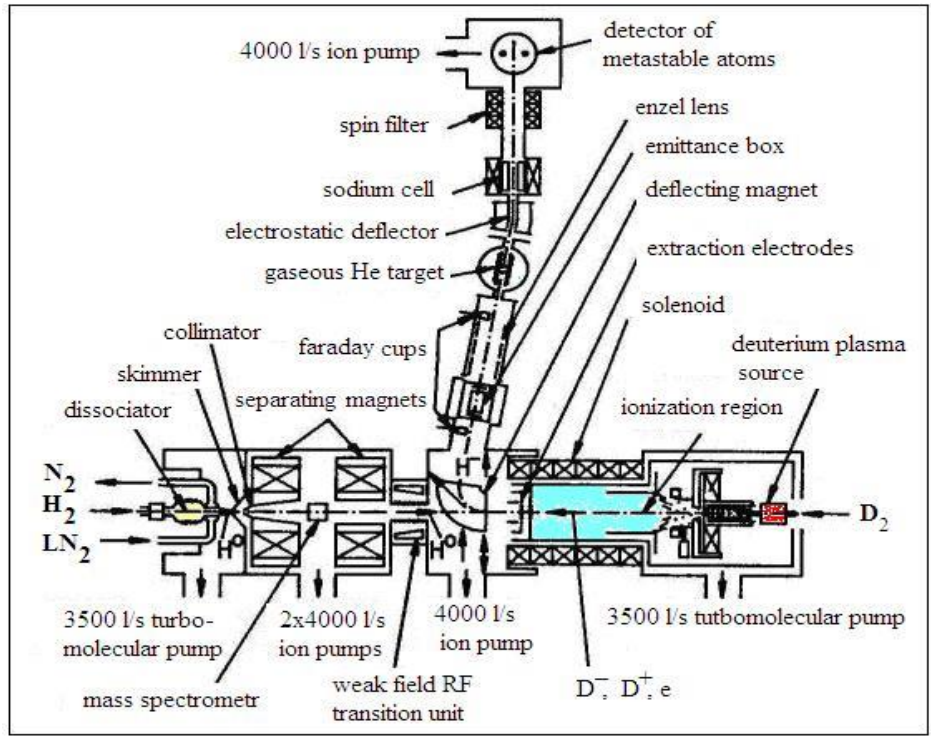

**FIGURE 1**: Schematic diagram of an ABPIS with a resonant charge exchange ionization [5].

1. Source of a polarized atomic H or D beam (left).
2. Surface-plasma source of cold unpolarized negative D$^-$ or H$^-$ ions with an arc discharge plasma source and a surface-plasma ionizer with cesium catalysis (right).
3. Charge exchange solenoid with a grid extraction system (middle).

4. Deflecting magnet for separation of the polarized and unpolarized H⁻ and D⁻ beams (transition to the top part).
5. Beam line and polarimeter (top part). The online polarization measurement is very important for optimization of many parameters influencing the polarization.

Several schemes of sextupole magnets and RF transition units are used in hydrogen or deuterium ABPIS. For atomic hydrogen, a typical scheme consists of two sextupole magnets followed by weak-field and strong-field RF transition units. In this case, the theoretical proton polarization will reach $P_z = 1$. Switching between the two $P_z = \pm 1$ states is performed by switching between the operation of the weak-field and strong-field RF transition units. For atomic deuterium, two sextupole magnets and three RF transitions are used in order to get deuterons with vector polarization of $P_z = 1$ and tensor polarization of $P_{zz} = 1, -2$. The polarized atomic beam intensity is proportional to the solid angle $\Delta\Omega = \pi\alpha^2$, which is determined by the magnetic focusing system and magnetic field:

$$\Delta\Omega = \pi\alpha^2 = \pi\mu B/\kappa T$$

$$B = 1.6\ T;\ \Delta\Omega = 1.5 \cdot 10^{-2}\ sr;\ \alpha = 0.07\ rad$$

$$B = 4.8\ T;\ \Delta\Omega = 4.5 \cdot 10^{-2}\ sr;\ \alpha = 0.21\ rad$$

Different methods for ionizing polarized atoms and their conversion into negative ions were developed in many laboratories. The techniques depended on the type of accelerator where the source was used and the required characteristics of the polarized ion beam (see ref. [3] for a review of the existing sources). For a pulsed ABPIS, the most efficient method was developed at INR, Moscow [5-9]. Polarized hydrogen atoms at thermal energy are injected into an ionizer solenoid with an incident flux of deuterium plasma where polarized protons or negative hydrogen ions are formed due to the quasi-resonant charge-exchange reactions:

$$H^0\uparrow + D^+ \Rightarrow H^+\uparrow + D^0$$
$$D^0\uparrow + H^+ \Rightarrow D^+\uparrow + H^0$$
$$H^0\uparrow + D^- \Rightarrow H^-\uparrow + D^0$$
$$D^0\uparrow + H^- \Rightarrow D^-\uparrow + H^0$$
$$^3He^0\uparrow + {}^4He^+ \Rightarrow {}^3He^+\uparrow + {}^4He^0$$
$$^3He^0\uparrow + {}^4He^{++} \Rightarrow {}^3He^{++}\uparrow + {}^4He^0$$

In an ABPIS, a $D^+$ plasma jet emerging from an arc discharge source is converted into a low-energy $D^-$ ion jet in a surface plasma ionizer. The design of a Surface Plasma (SP) ionizer (INR version) is shown in Fig. 3.

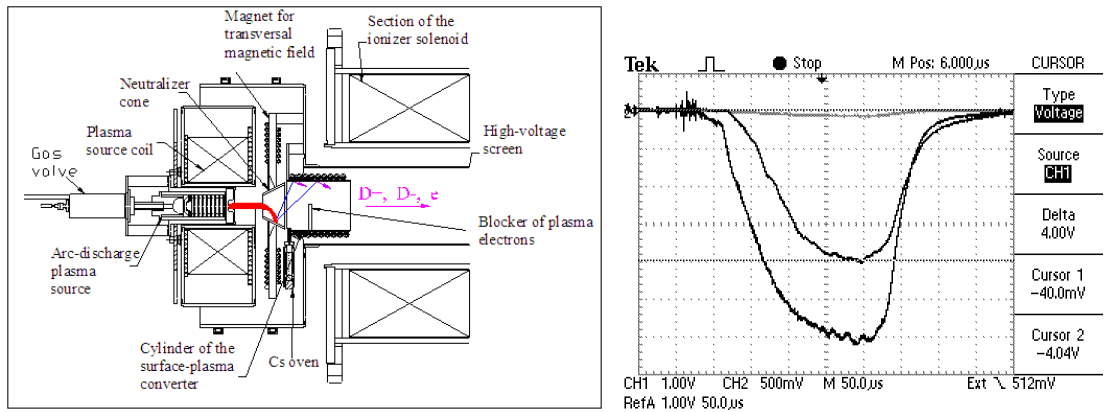

**FIGURE 2**: Generation of a cold D⁻ plasma jet for resonant charge exchange negative ionization of polarized hydrogen atoms (INR version) [3].

**FIGURE 3:** Oscillograms of a polarized H⁻ ion current of up to 4 mA (the vertical scale is 1 mA/div) and of an unpolarized D⁻ ion current of up to 60 mA (10 mA/div) in INR ABPIS [3].

By using resonant charge exchange ionization, it is possible to have an efficiency of the polarized atom to polarized H⁻ transformation of over 12%. The high selectivity of the polarized atom ionization allows polarization values above 0.9. The arc-discharge plasma source developed at BINP [10] is used as a plasma jet generator. A converter with cesium deposition is used for conversion of the plasma flux into negative ions. The first cesiated converter for negative ion generation was proposed by V. Dudnikov [5] and was further optimized by A. Belov [3].

The development and adaptation of a high-polarization ABPIS promises to improve the productivity of very costly polarized colliding beam experiments at RHIC and is crucial for the electron ion colliders under development at JLab and BNL. Polarized H⁻/D⁻ currents were increased up to 4 mA with measured polarization P of up to 0.95. Unpolarized D⁻/H⁻ currents were 60/100 mA. The coextracted electron current was efficiently suppressed by a blocker.

The signals of polarized and unpolarized beams from the INR ABPIS are shown in Fig. 3 [3]. The pulsed polarized negative ion source (CIPIOS) produced multi-mA beams for injection into the Indiana Cooler Injector Synchrotron (CIS) under regular operation for several years [8,9]. A schematic of the ion source and LEBT is shown in Fig. 4. Parameters of the existing polarized sources are listed in Table 1.

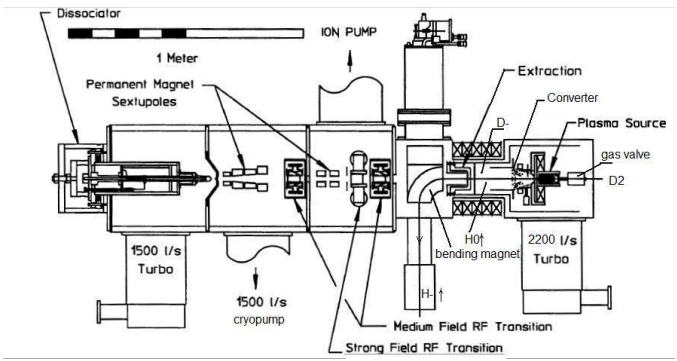

**FIGURE 4:** Design of the CIPIOS-CIS ABPIS [8,9]. The beam is extracted from the ionizer toward the ABS and is then deflected downward with a magnetic bend and towards the RFQ with an electrostatic bend. This results in a nearly vertical polarization at the RFQ entrance [9].

Table 1: Existing source parameters

| OPPIS/BNL, H⁻ only | Pulse Width | 500 µs (up to DC?) |
|---|---|---|
| (In operation) | Peak Intensity | 1.6 mA |
|  | Max Pz | 86% of nominal |
|  | Emittance (90%) | 2.0 π·mm·mrad |
| IUCF/INR CIPIOS: | Pulse Width | Up to 500 µs |
| (Shutdown 8/02) | Peak Intensity H⁻/D⁻ | 2.0 mA/2.2 mA |
|  | Max Pz/Pzz | 85% to > 90% |
|  | Emittance (90%) | 1.2 π·mm·mrad |
| INR Moscow: | Pulse Width | > 100 µs |
| (Test Bed Only) | Peak Intensity H⁺/H⁻ | 11 mA/4 mA |
|  | Max Pz | 80%/95% |

| | Emittance (90)% | 1.0 π·mm·mrad/ 1.8 π·mm·mrad |
|---|---|---|

# PROPOSED MODIFICATIONS OF A UNIVERSAL ABPIS

For polarized proton and deuteron accumulation in the EIC, we propose to use an atomic beam polarized ion source (ABPIS) of negative $H^-/D^-$ ions. This version of the source can be used with minor modifications for production of polarized and unpolarized $H^-$, $D^-$, $T^-$, $H^+$, $D^+$.

The highest possible polarization is required to reduce the systematic and statistical errors in polarization experiments. The double spin asymmetry statistical error is proportional to ~ $1/\sqrt{L P^4}$; therefore, a 5% polarization increase at the source (or 5% reduction of the polarization loss during acceleration, for instance, in the AGS and RHIC) is effectively equivalent to a 30% increase in the data taking time [4]. By using charge-exchange injection and special equipment for polarization preservation (e.g. figure-8 rings) [1], it is possible to increase the accelerated beam intensity and essentially eliminate the polarization loss but impossible to enhance the beam polarization of heavy particles once they leave the source.

In these ABPIS, polarized $D^-$ ions are produced in charge exchange of polarized $D^0$ with unpolarized $H^-$. $D^-$ polarization is diluted by formation of unpolarized $H^-$ and $D^-$ in the H plasma interaction with the surface of the converter. These unpolarized $H^-$ ions are produced from an admixture of $H_2$ gas in the $D_2$ gas, and from the hydrogen and water adsorbed in the gas delivery system of the arc discharge plasma source and on the ionizer surfaces.

We propose to increase the beam particle polarization up to the highest level by optimization of ABPIS components and suppression of parasitic depolarizing processes. The ABPIS of INR with the highest polarized beam parameters [3] was built nearly 30 years ago using the technology and components available at that time. Much more advanced components with better characteristics and better reliability are available now and will be developed in this project. These new components can be used to develop a novel and significantly improved ABPIS design. To date, there have been some large, expensive, multi-year efforts involving international experts in the development of polarized atomic beams for polarized target and polarimetry projects with a clear scientific payoff [11]. We can use these developments to improve our ABPIS design.

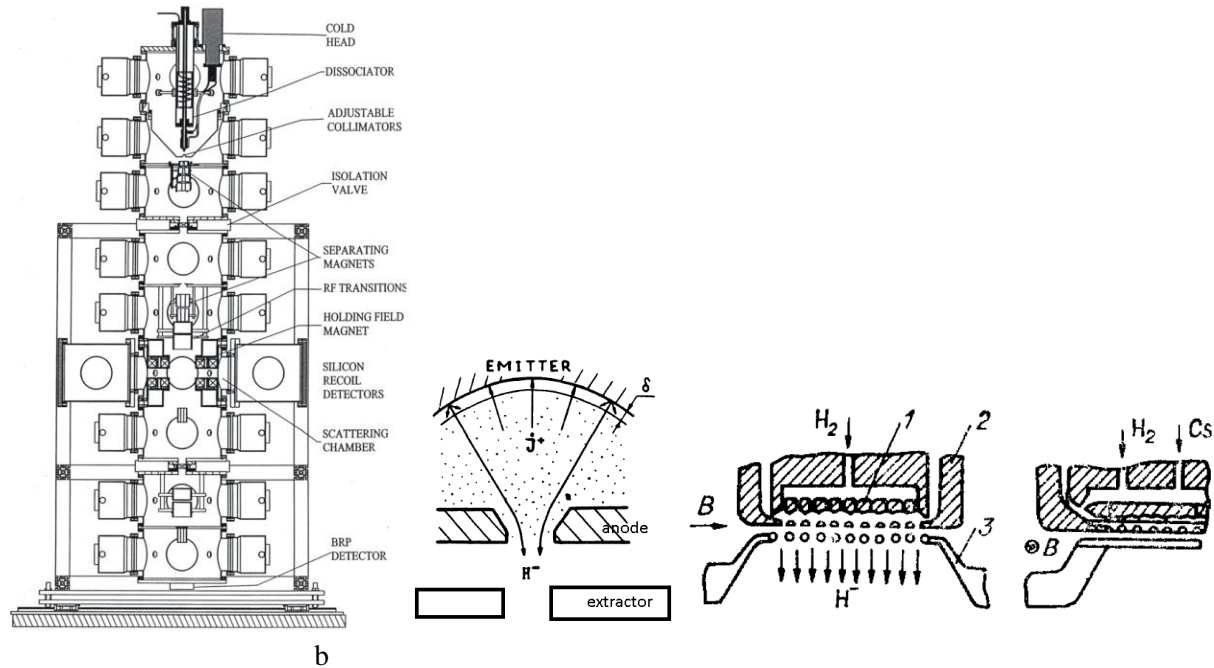

b

**FIGURE 5**: Schematic of the BNL ABPS; the H-jet polarimeter includes three major parts: polarized Atomic Beam Source (ABS), scattering chamber, and Breit-Rabi polarimeter.
**FIGURE 6**. Cylindrical or spherical surface of emitter electrode is focused emitted negative ion into small slit or round hole.
**FIGURE 7.** Schematic of a multispherica semiplanotron.

For the new ABPIS design, we will use the general design of the CIPIOS-CIS ABPIS [8,9] shown in Fig. 4 as a prototype. We will improve on it by using novel design concepts and the most advanced components including 1) a superconducting or strong permanent sextupole magnet; 2) a new advanced RF dissociator with helicon discharge in a magnetic field; 3) new design of surface plasma ionizer with multi spherical focusing, 4) an optimized high-transparency heated extraction system, and 5) an optimized low-aberration bending magnet. A superconducting sextupole will be used for the highest intensity polarized beam production. An advantage of using a strong permanent sextupole magnet is that it has a lower cost of manufacturing and operation while still providing the highest polarization. The existence and availability of the components for the ABPIS construction is a great advantage of this project. Many of the necessary components and materials are now available with much better parameters.

The BNL polarized atomic H jet [11] has an intensity of up to $1.7 \cdot 10^{17}$ atoms/s and polarization of ~0.97 in the DC mode of operation and it is possible to have an even higher intensity and polarization in the pulsed mode of operation. An ABS with very good parameters was developed for the internal target experiments at BINP [12]. This ABS uses a superconducting sextupole focusing system with magnetic field at the pole tip of up to 4.8 T. Sextupole focusing systems with permanent magnets have a magnetic field at the pole tip of up to 1.6 T. A schematic of the BNL ABPS is shown in Fig. 5. [13]

The polarimeter axis is vertical and the recoil protons are detected in the horizontal plane. The common vacuum system is assembled from nine identical vacuum chambers, which provide nine stages of differential pumping. The system building block is a cylindrical vacuum chamber 50 cm in diameter and of 32 cm length with four 20 cm (8.0") ID pumping ports. There are 19 TMPs with 1000 l/s pumping speed for hydrogen.

Efficiency of negative ion generation were improved significantly by invention geometrical focusing. Cylindrical or spherical surface of emitter electrode is focused emitted negative ion into small slit or round hole, as shown in Fig. 7.

A multispherica semiplanotron was developed in [14] for high intense H- beam production as shown in Fig. 7. A photograph of the multispherica semiplanotron is shown in Fig. 8.

Suppression a gas flow from SPS by discharge is show in Fig. 9. Here $q_o$ is gas flux without discharge, $q_p$ is gas flow with discharge, $I_d$ is a discharge current, $\delta$ is cross talk signal. It is possible to decrease the gas flux from SPS up to 5 time and increase a gas efficiency up to 30%.

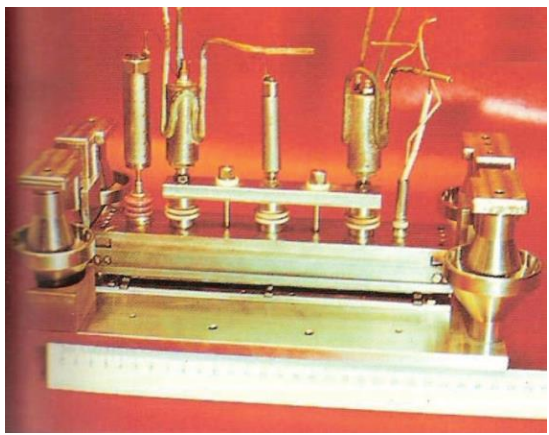
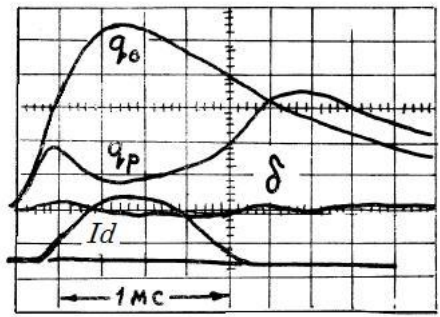

**FIGURE 8**. A photograph of the multispherica semiplanotron.
**FIGURE 9**. Suppression a gas flow from SPS by discharge.

The H- beam intensity from this SPS was increased up to 12 A. The ionizer is shown schematically in Figure 10. Its operation is basically the same a normal magnetron surface plasma source, although with an inverted geometry, i.e, the cathode is the outer and the anode the inner of two concentric cylinders. H- ions are produced on the Cs-coated molybdenum cathode with spherical concaves, and are accelerated away from the cathode. through small holes in the anode, and to the central region of the ionizer.

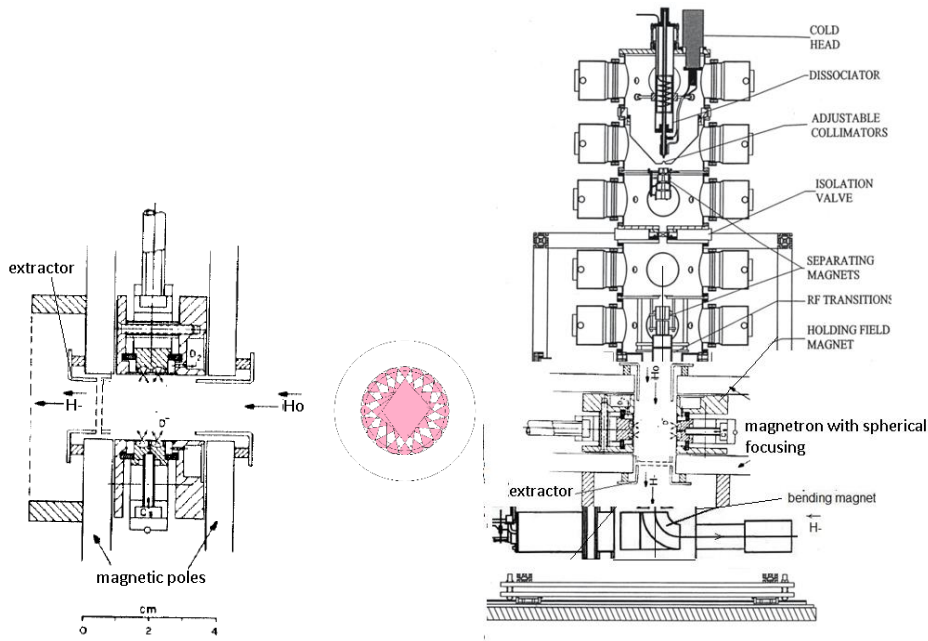

**FIGURE 10**. . Resonance charge exchange ionizer with a spherical focusing for ionization of polarized Ho or Do.
**FIGURE 11.** High polarization negative ion source with surface plasma ionizer with spherical focusing

Enough H+ Ions from the source plasma diffuse into the central region to provide space charge neutralization of the H- and D-. Polarized Do passes axially through the central region and is ionized. The short length of this ionizer, compared to other techniques, gives it the advantagec of having a large acceptance for the polarized atomic beam. A spherical geometrical focusing of produced negative ions is increase gas efficiency and energy efficiency of ionizer. Proposed high polarization negative ion source with surface plasma ionizer with spherical focusing is shown in Fig. 11. It consist of source of polarized atoms as BNL polarimeter with multispherical focusing surface plasma ionizer, extraction grids and bending magnet for separation polarized D- and non-polarized H-.

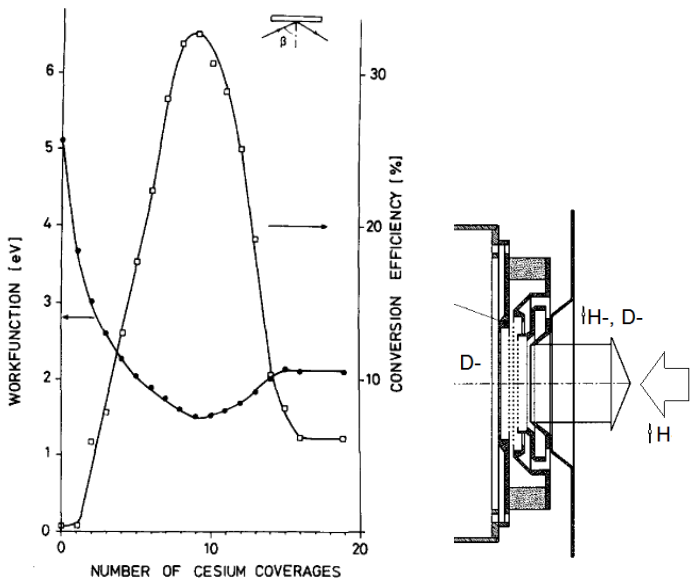
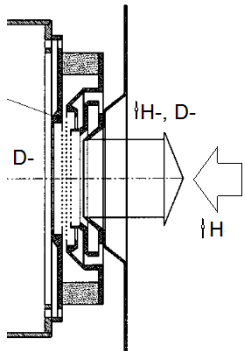

**FIGURE 12**: Conversion efficiency of $H^+$, $H^0$ and surface work function vs cesium surface concentration.
**FIGURE 13**: Transparent multigrid extraction system.

The probability of negative ion emission from a cesiated surface bombarded by plasma ions and atoms strongly depends on the surface work function (WF) as shown in Fig. 12. It is very important to keep the surface WF as low as possible. Special activation procedures were recently developed for production of efficient long-term stable cesiation in SPS [15] and can be reproduced in ABPIS.

A very transparent multigrid extraction system shown in Fig. 13 can be used for improved beam formation and emittance decrease. A four-electrode multislit extraction consists of three multi-wire grids and a fourth cylindrical grounded electrode. The grids are made of 0.2 mm molybdenum wire. The spacing between wires is 1.0 mm. The wires are positioned on the mounting electrodes by precisely cut grooves and fastened by point welding.